\documentclass[twoside]{article}
\usepackage[latin1]{inputenc}
\usepackage{t1enc}
\usepackage{a4}
\usepackage{tabularx}
\usepackage{epsfig}

\def\ref{\vspace{4pt}\noindent\hangindent=10mm}
\def\spose#1{\hbox to 0pt{#1\hss}}
\def\lta{\mathrel{\spose{\lower 3pt\hbox{$\mathchar"218$}}
     \raise 2.0pt\hbox{$\mathchar"13C$}}}
\def\gta{\mathrel{\spose{\lower 3pt\hbox{$\mathchar"218$}}
     \raise 2.0pt\hbox{$\mathchar"13E$}}}
\def\etal{{\it et al.\ }}
\def\R{${\cal R}$}

\begin{document}

\setcounter{figure}{0}
\setcounter{section}{0}
\setcounter{equation}{0}

\begin{center}
{\Large\bf
The Evolution of Galaxies \\[0.2cm]
on \\[0.5cm]
Cosmological Timescales} \\[0.7cm]

Uta Fritze -- v. Alvensleben \\[0.17cm]
Universit\"atssternwarte G\"ottingen \\
Geismarlandstr. 11, D -- 37083 G\"ottingen, Germany \\
ufritze@uni-sw.gwdg.de
\end{center}

\vspace{0.5cm}

\begin{abstract}
\noindent{\it
I discuss the chemical and spectrophotometric evolution of galaxies over cosmological timescales and present a first attempt to treat both aspects in a chemically consistent way. In our evolutionary synthesis approach, we account for the increasing metallicity of successive generations of stars and use sets of stellar evolutionary tracks, stellar yields, spectra, etc. for various metallicities. This gives a more realistic description of nearby galaxies, which are observed to have broad stellar metalllicty distributions, as well as of young galaxies at high redshift. Selected results are presented for the chemo-cosmological evolution of galaxies as compared to QSO absorption line observations and for the spectrophotometric evolution of galaxies to very high redshifts. On cosmological timescales, intercations are important drivers of galaxy evolution. Neglecting dynamical aspects we study the effects of interaction-induced starbursts on the spectrophotometric and chemical evolution of galaxies and briefly discuss the formation of star clusters and Tidal Dwarf Galaxies in this context.

}
\end{abstract}

\section{Introduction}
The evolution of galaxies on cosmological timescales has 3 aspects: the chemical, spectral, and dynamical evolution. In reality, all of them are intimately coupled  as shown by the existence of several observational relations involving quantities from any two of these 3 domains. Examples are the Faber -- Jackson, Fundamental Plane, or luminosity -- metallicity relations for ellipticals or the 
Tully -- Fisher relation and trends of characteristic HII region abundances, average colours, emission line strengths, luminosities, and mass - to - light ratios with spiral type, i.e. with the transition from bulge-dominated to disk-dominated systems.

While over short timescales or lookback times and, at least, for giant galaxies, the 3 aspects of galaxy evolution might be treated independently, any modelling of galaxy evolution over cosmological timescales definitely requires a consistent treatment of all 3 aspects. Unfortunately, this is still too complex today, even for numerical simulations on powerful machines. 

Independently, all 3 aspects of galaxy evolution modelling have quite some tradition already. We are, however, still at the very beginning of a consistent understanding of galaxy formation and evolution over cosmological timescales including all 3 aspects. 
A consistent modelling of the chemical and dynamical evolution of galaxies is e.g. attempted by the group of G. Hensler in Kiel (`Chemodynamics'). 
With the availability of stellar input physics for various metallicities, we developed in G\"ottingen a consistent description of the chemical and spectral evolution over cosmological timescales, i.e. coupled to a cosmological model. This {\bf chemically consistent evolutionary synthesis} method and selected applications for the interpretation of galaxy observations over a wide redshift range will be presented here.

While still being far from a consistent coupling of our chemically consistent chemical, spectrophotometric and cosmological model with a dynamical galaxy evolution model, we at least started to study the starbursts accompanying strong interactions and mergers involving gas rich galaxies and their impact on the chemical and spectral evolution of galaxies. This brings us to some surprising features accompanying those strong bursts: the formation and evolution  both of a new population of bright and massive young star clusters and of a new class of dwarf galaxies forming from recycled material along the tidal features of the merging giant galaxy pair (Tidal Dwarf Galaxies).

\section{Chemically Consistent Evolutionary Synthesis}
\subsection{Observational Evidence for Broad Metallicity Distributions and Subsolar Metallicities}
Evidence for extended stellar metallicity distributions and, in particular, for the importance of subsolar abundance contributions has been accumulating over the last years. Stars in elliptical galaxies and bulges, e.g., show a metallicity distribution extending over more than a factor 10 from ${\rm 0.06 \leq Z_{\ast}/Z_{\odot} \leq 2.5}$ with average metallicity slightly below solar (Mc William \& Rich 1994), G dwarf stars in the solar neighbourhood cover a range from ${\rm 0.15 \leq  Z_{\ast}/Z_{\odot} \leq 1.5}$ with, again, the average metallicity being subsolar (Rocha-Pinto \& Maciel 1996). Characteristic HII region abundances (i.e. measured at 1 ${\rm R_e}$) in spirals are observed to be in the range from ${\rm \sim Z_{\odot}}$ for early type spirals Sa/b through ${\rm \ll Z_{\odot}}$ for late type spirals Sd (Oey \etal 1993, Zaritsky \etal 1994, Ferguson \etal 1998). Dwarf galaxies of all classes (dEs, dSphs, dIs, BCDs) show -- sometimes substantially -- subsolar metallicities, both in their stellar populations and in their gas phase abundances (e.g. Richer \& McCall 1995). So, while already in the local Universe, subsolar abundances are clearly prevailing on global galaxy scales -- with the exception of some central regions of massive luminous galaxies -- this is, of course, even more so the case in the early Universe.  

From restframe UV stellar wind lines, redshifted into the optical for star forming galaxies at redshifts 3 -- 4, stellar abundances around 1/10 ${\rm Z_{\odot}}$ are derived (cf. Lowenthal \etal 
1997, Trager \etal 1997). Damped Ly$\alpha$ absorbers -- most probably the progenitors of present day galactic disks -- feature ISM abundances in the range ${\rm 10^{-3} < Z_{ISM}/Z_{\odot} < 1}$ over a redshift range ${\rm 0.4 \lta z \lta 4.5}$ (e.g. Pettini \etal 1997, 1999, see also Sect. 5). 

These are only a few selected examples from a much longer and still increasing list of direct observational hints to the existence of significant metallicity distributions in galaxies and to the importance of subsolar metallicities. 

\subsection{Method and Input Physics}

From a very basic consideration it is immediately clear that any stellar system with star formation ({\bf SF}) extending over more than the lifetime of the most massive stars ($10^6$ yr) is composite both in terms of age {\bf and} metallicity.

While star clusters are {\bf S}imple {\bf S}tellar {\bf P}opulations ({\bf SSP}s) consisting of one stellar generation with one age and one metallicity, any galaxy is a {\bf C}omposite {\bf S}tellar {\bf P}opulation ({\bf CSP}) with its stars spanning finite ranges both in age and metallicity. 

The {\bf G\"ottingen Evolutionary Synthesis Model} starts from a gas cloud of given mass, assumes some  {\bf S}tar {\bf F}ormation {\bf H}istory ({\bf SFH}) and {\bf I}nitial {\bf M}ass {\bf F}unction ({\bf IMF})  -- the basic parameters of this kind of approach -- and then, with some book-keeping algorithm, calculates the time evolution of the stellar population across the HRD from a set of stellar evolutionary tracks. With photometric calibrations for colours or absorption indices along all stellar evolutionary tracks the photometric evolution in terms of luminosities, colours, and absorption indices is obtained. Assigning stellar spectra to any point along the tracks and weighting them with the numbers of stars calculated at these points at any given time yields the time evolution of a synthetic galaxy spectrum. Integrating over the synthetic galaxy spectrum with response function for any filter system (e.g. Johnson UBVRIJHK, HST F300W, F450W, ..., F814W), luminosities in the respective passbands are obtained in the same way as for an observed galaxy spectrum. 

In this approach, evolutionary consistency is guaranteed and the time evolution can explicitly be studied. This opens the possibility to directly couple to a cosmological model and study the spectrophotometric evolution as a function of redshift over cosmological timescales (cf. F.-v.A. 1989). 

Evolutionary synthesis modelling {\sl per se} accounts for the age distribution of the stars within a galaxy and its time evolution (for a review on synthesis methods see F.-v.A. 1994). To properly account for the metallicity distributions in composite systems like galaxies, the abundance evolution in 
the gas has to be followed in order to describe successive generations of stars forming out of this gas with stellar evolutionary tracks, yields, and spectra appropriate for their respective initial metallicities, i.e. the gas metallicity at their birth. We call this approach {\bf chemically consistent} (= {\bf cc}). 

A modified form of Tinsley's equations (Tinsley 1980) with stellar yields for SNII, SNI, PN, and stellar mass loss  is solved to obtain the gas content G, the global metallicity Z of the ISM, as well as individual element abundances [${\rm X_i/H}$], and abundance ratios [${\rm X_i/X_j}$]. 
We use sets of input physics 
for 5 different stellar metallicities in the range 
\begin{center} ${\rm  -2.3~ \leq ~[Fe/H]~ \leq ~+0.3}$ \end{center} 
with logarithmic element abundances with respect to solar defined by [X$_{\rm i}$/H] $:=~ log {\rm (X_i / X_H)} ~-~ log {\rm (X_i^{\odot} / X_H^{\odot}}$).
  These sets of input physics for the description of both the spectral {\bf and} chemical evolution of model galaxies comprise 
stellar evolutionary tracks -- covering all relevant evolutionary phases from ZAMS through the PN phase or SN explosion -- lifetimes, and remnant masses from the Padova group (Bressan \etal 1993, Fagotto \etal 1994a, b, c). Evolution of low mass stars is taken from Chabrier \& Baraffe 1997. 
Stellar yields for a series of individual elements from $^{12}$C through $^{56}$Fe for massive stars (${\rm >8~M_{\odot}}$) are from Woosley \& Weaver 1995 and 
yields for intermediate mass stars from van den Hoek \& Groenewegen 1997. 
SNIa contributions 
to Fe, C, ..., are included for the carbon deflagration white dwarf binary scenario as outlined 
by Matteucci \& Greggio 1986. 
We caution that 
metallicity dependent stellar yields depend on 
${\rm \frac{\Delta Y}{\Delta Z}}$, explosion energies, remnant masses, etc. 
and the metallicity dependence of these factors is still poorly understood. 
SNIa yields are only available for ${\rm Z_{\odot}}$ (Nomoto 
\etal 1997, model W7). No important metallicity dependence is expected for SNIa yields 
except for a possible lower metallicity limit to the explosion (Kobayashi \etal 1998). 
Model atmosphere spectra, colour and absorption index calibrations are from Lejeune \etal 1997, 1998, and Worthey \etal 1994, respectively. 

Although this cc approach goes significantly beyond what was possible before, it still is a simplification in the sense that all the input physics is only available for scaled {\bf solar abundance ratios}. Abundance ratios in galaxies are determined by the metallicity dependent stellar yields, the SFH and IMF, and, in general, will be non-solar. I.e., 
stellar evolution and galaxy evolution themselves are implicitly coupled and the coupling depends on the SFH and IMF (cf. F.-v.A. 1998b). 
For a review on chemically consistent evolutionary synthesis see F.-v.A. 1999b. 

The spectrophotometric evolution of the stellar component of galaxies and 
the chemical evolution of ISM abundances both are studied not only as a function of time, but also 
-- for any cosmological model as given by ${\rm H_o, \Omega_o, \Lambda_o,}$ and a redshift of 
galaxy formation ${\rm z_f}$ -- as a function of redshift.

\subsection{Model Parameters and Local Templates}
We use an IMF in the form given by Scalo 1986 with lower and upper mass limits of 0.08 and 85 ${\rm M_{\odot}}$, respectively, and SFHs, ${\rm \Psi(t)}$, appropriate for the various spectral types of galaxies. 
For ellipticals we use a standard 
${\rm \Psi(t) \sim e^{-t/t_{\ast}}}$. For spiral types Sa ... Sc the {\bf S}tar {\bf F}ormation {\bf R}ate ({\bf SFR}) at any timestep is tied to the gas-to-total 
mass ratio, ${\rm \Psi(t) \sim \frac{G}{M} (t)}$, and for Sd galaxies ${\rm \Psi(t) = } ~const.~$ is assumed. 
Characteristic timescales for SF ${\rm t_{\ast}}$ (for spirals defined via 
${\rm \int_0^{t_{\ast}} \Psi \cdot dt = 0.63 \cdot G\mid_{t=0}}$) thus range from 1 Gyr for ellipticals 
to 2, 3, 10, and 16 Gyr for Sa, Sb, Sc, and Sd galaxies, respectively. 

The SFHs, together with the IMF, have been chosen as to provide agreement 
of our model galaxies 
after a Hubble time of evolution with integrated colours, luminosities, absorption features (E/S0s), 
emission line strengths (spirals), and template spectra from the UV through NIR (Kennicutt 1992, Kinney \etal 1996, cf. M\"oller \etal 1998), as well as with 
characteristic 
HII region abundances ($:=$ measured at the effective radius) typical for the respective galaxy types (Zaritsky \etal 1994, Oey \& Kennicutt 1993, van Zee \etal 1998, Ferguson \etal 1998). 

\subsection{Advantages and Shortcomings}

The combined approach to study the chemical evolution of ISM abundances {\bf and} the spectrophotometric evolution of the stellar population allows -- with the same number of parameters (IMF and SFH) as any single aspect model -- for a much larger number of model observables (spectra, luminosities and colours from UV through NIR, emission and absorption line strengths, 
gas content and metallicity) to be compared to observations. The combination with a cosmological model provides a long redshift or time baseline for comparison with galaxy data, 
allowing not only to test the models, constrain the parameters, but also to make numerous predictions testable by future observations (F.-v.A. PhD Thesis 1989). The G\"ottingen Evolutionary Synthesis code has a remarkable analytical potential. It is possible to trace back -- at any time and in its time or redshift evolution -- the luminosity contribution to any wavelength band of every single stellar mass, of the various spectral types, luminosity classes, and metallicity subpopulations. 
Ejection rates of every individual element, as well, are monitored for every stellar mass, nucleosynthetic origin (PN, SNII, SNIa), and metallicity subpopulation. 

The SFHs we use for different spectral types of galaxies are similar to those used by other groups (Bruzual \& Charlot 1993, Rocca -- Volmerange \& Guiderdoni 1988) and meant to be global SFRs -- averaged over the entire galaxy and reasonable intervals of time. SFRs fluctuating around our smooth SFHs -- either locally or on short timescales -- could not be discriminated from their smooth idealisations in integrated galaxy properties after sufficiently long lookback times. 

Our models are simple 1-zone descriptions without any
dynamics or spatial resolution, meant to describe global average
quantities like integrated spectra, luminosities, colours, emission and absorption line 
strengths, or characteristic HII region abundances as measured around the effective radius. 
While the finite lifetimes of individual stars before they restore their partly enriched material to the ISM are properly accounted for in the chemical modelling (we do NOT use an Instantaneous Recycling Approximation), the mixing of the gas is assumed to happen instantaneously. 
If not indicated otherwise, 
models are closed boxes without any inflow (or outflow). While, clearly, real galaxies may not be closed boxes 
over cosmological timescales, this simplification reflects our ignorance of the time or redshift 
evolution of gas infall rates and infall abundances. Those cannot yet be taken from 
hierarchical galaxy formation models that are restricted to Dark Matter. The only consistent way to account for mass (and energy) exchange beween a (proto-)galaxy and its environment seems to be a coupling of our chemo-spectrophotometric evolution with models for cosmological structure and galaxy formation (see Contardo, F.-v. A. \& Steinmetz 1998 for a first attempt). 

\subsection{Results for Nearby Galaxies}
Our cc spectrophotometric evolution models were first presented in M\"oller \etal 1997 where we discuss the comparison 
between the (mass-weighted) ISM
metallicities  for various galaxy types and the luminosity-weighted
metallicities of their stellar populations as seen in different wavelength
bands. At late stages, stars in models with $const.$ SFR (Sd) show a
stellar metallicity distribution strongly peaked at ${\rm \frac{1}{2}
Z_{\odot}}$ at all wavelengths and close to the ISM
metallicity. Elliptical models, in agreement with observations, show broad stellar
metallicity distributions extending from ${\rm Z=10^{-4}~to~Z=0.05}$
in all bands (cf. Fig. 1 in F.-v. A. 1999b). 

In collaboration with D. Calzetti (STScI) we are currently working on a consistent inclusion of dust, tying the amount of dust to the evolution of both the gas content and the metallicity in our models, as well as including different spatial distributions of dust and stars in different galaxy types (cf. M\"oller \etal {\sl astro-ph/9906328} for first results).

\section{CC Spectro -- Cosmological Evolution: \\
Selected Results}
Before the spectrophotometric evolution as a function of redshift can be studied for any given set of cosmological parameters, evolutionary corrections ${\rm e_{\lambda}}$ (since a galaxy 
at ${\rm z > 0}$ is younger) and cosmological corrections -- also called k-corrections -- ${\rm k_{\lambda}}$ (since its spectrum 
is redshifted) are required. 

For redshifts ${\rm z \gg 1}$ the attenuation of galaxy light by the cumulative effect of 
intervening hydrogen has to be taken into account. 
Intergalactic neutral hydrogen HI largely comes in the form of Ly$\alpha$ clouds causing a forest of narrow low column density absorption lines in the featureless continua of background QSOs. The cumulative effect of intergalactic HI distributed stochastically along the line of sight to galaxies at ${\rm z \gta 2.5}$ has been shown by Madau (1995) to significantly attenuate the spectra of distant galaxies at restframe wavelengths below 1216 \AA. The average attenuation obtained from a large sample of sightlines given by Madau \etal 1996 is applied to our redshifted model galaxy spectra. It additionally weakens their fluxes below restframe ${\rm \lambda = 1216}$ \AA \ and its effect is included in our cosmological corrections.  

The age of a galaxy at redshift ${\rm z = 0}$ is given by 

\begin{center}  ${\rm t_o := t_{gal}(z=0) := t_{Hubble}(z=0) - t_{Hubble}(z_f)}$ \end{center}

The red colours of present-day elliptical galaxies require galaxy ages in the range 12 -- 15 Gyr. Reasonable combinations of the cosmological parameters are therefore restricted by this minimum age requirement. 

Model galaxy spectra convolved with filter response functions yield absolute fluxes and magnitudes. Luminosities at ${\rm z = 0}$ are normalised to the average absolute B - band luminosities of the respective galaxy types observed in Virgo (cf. Sandage \etal 1985) before redshifting the synthetic spectra.  

Apparent magnitudes ${\rm m_{\lambda}}$ of high redshift galaxies in any filter ${\rm \lambda}$ are obtained from model absolute magnitudes ${\rm M_{\lambda}}$ via 
the bolometric distance modulus BDM(${\rm H_o,~\Omega_o,~\Lambda_o)}$, evolutionary and cosmological corrections:
\begin{center} ${\rm m_{\lambda}(z) = M_{\lambda}(z=0,~t_o) + BDM(z) + e_{\lambda}(z) + k_{\lambda}(z) }$ \end{center}
The cosmological or k - correction ${\rm k_{\lambda}}$ in any wavelength band $\lambda$ accounts for the magnitude difference between a galaxy of age ${\rm t_o}$ locally and the same galaxy spectrum redshifted to z
\begin{center} ${\rm k_{\lambda}(z) := M_{\lambda}(z,~t_o) - M_{\lambda}(0,~t_o)}$ \end{center} 
and can also be calculated for observed galaxy spectra. In this case, the maximum redshift to which this is possible depends on how far into the UV the observed spectrum extends. Our model galaxy spectra at ${\rm z=0}$ extend from 90 \AA \ through 160 $\mu$m and, hence, allow for 
cosmological corrections in optical bands up to ${\rm z \gg 10}$. 
Evolutionary corrections ${\rm e_{\lambda}}$ account for the age difference between a galaxy of today's age ${\rm t_o}$ and the same galaxy at a younger age ${\rm t_{gal}(z)}$ corresponding to the redshift z when its light was emitted
\begin{center} ${\rm e_{\lambda}(z) := M_{\lambda}(z,t_{gal}(z)) - M_{\lambda}(z,~t_o)}$ \end{center}
Evolutionary corrections, of course, cannot be given without an evolutionary synthesis model. 
It is important to stress that both the evolutionary and the cosmological corrections do not only depend on the cosmological parameters but also on the SFH or spectral type of the galaxy. 

In M\"oller \etal 2000a ({\sl in prep.}) the redshift evolution of cc galaxy spectra, evolutionary and cosmological corrections, apparent magnitudes, and colours will be presented for two different cosmologies.

\subsection{CC Models Compared to Solar Metallicity Models} 
A first comparison of the spectrophotometric evolution of model galaxies described in a cc way with those described using only solar metallicity input physics is presented in M\"oller \etal 1997. 

The most important result is that cc spiral models, as compared to ${\rm Z_{\odot}}$ models, are brighter by 1 -- 2 mag in B and ${\rm \lta 1.5}$ mag in K at redshifts ${\rm z \gta 1}$. 

This has significant implications for the interpretation of high redshift galaxy data, e.g. for our understanding of the Faint Blue Galaxy Excess. It means that -- contrary to earlier expectations -- normal spiral galaxies at ${\rm z \sim 1~.~.~.~\gta 2}$ can still contribute to galaxy counts around ${\rm B \sim 27}$ mag, and if counts are compared to cc models, less of an excess is expected. The impact of cc modelling on the Luminosity Functions ({\bf LF}s) in various wavelength bands and on the redshift evolution of LFs of specific galaxy types is currently being explored. 

\subsection{Comparing with Lyman Break Galaxies} 
In any type of galaxy at any time, the flux is close to zero below the Lyman break at 912 \AA. While galaxies with passively evolving stellar populations do not have important UV fluxes longward of 912 \AA \  either, those with active SF do show significant UV fluxes longward, while being self-absorbed shortward of $912$ \AA. At ${\rm z \gta 2.5}$ the Lyman break has moved beyond the U - band, causing star forming galaxies to drop out of deep U images while visible in B, V, etc. images. By 
${\rm z \gta 3.5}$ the Lyman break has moved beyond the V - band, causing star forming galaxies at ${\rm z \gta 3.5}$ to also drop out of the V - band while visible at longer wavelengths. {\sl Per definitionem} only galaxies with active SF are detected with this technique. A passive galaxy at redshift ${\rm z=3}$, if it existed, would neither be detected in U nor in V or any other optical band. Its restframe flux steeply increasing around 4000 \AA \ only, it would first appear in the NIR H-band. 

Specific colour criteria have been developed to isolate galaxy candidates at ${\rm 2.5 \leq z \leq 3.5}$ and at ${\rm 3.5 \leq z \leq 4.5}$, called {\bf L}yman {\bf B}reak {\bf G}alaxies ({\bf LBG}s) (see e.g. Steidel \etal 1995, 1999). 
Follow-up spectroscopy has proven the efficiency of this technique and provided fairly large samples of spectroscopically confirmed galaxies by today: $> 560$ galaxies at ${\rm z \sim 3}$ and $> 46$ galaxies at ${\rm z \sim 4}$ both from the {\bf H}ubble {\bf D}eep {\bf F}ield and ground based deep imaging surveys. The FORS Deep Field currently observed at the ESO-VLT will soon increase the number of very high redshift galaxies by a significant factor.  

The nature of the LBGs is controversial. Appearing quite compact with scale radii of 1 -- 3 kpc they were suspected to be the progenitors of present-day spheroidal galaxies or bulges (Steidel \etal 1996, Giavalisco \etal 1996, Fria\c ca \& Terlevich 1999) or subgalactic fragments (Somerville \etal 1998). 
Caution seems required, however, comparing rest frame UV scale lengths of LBGs with optical scale lengths of local galaxies, since local galaxies on UV images look very different in overall morphology -- not to mention scale lengths --  
from how they look in the optical (cf. Hibbard \& Vacca 1997).  
The high surface density of LBGs, their SFRs -- estimated from UV fluxes to be in the range 3 -- 60 ${\rm M_{\odot}~yr^{-1}}$ --, and luminosities are easier to understand if they are the progenitors of spirals. 

Comparing the redshift evolution of luminosities and colours of our cc galaxy models with the first set of spectroscopically confirmed HDF dropout galaxies (Lowenthal \etal 1997) in Fig. 1, we find that all these LBGs at ${\rm 2 \lta z \lta 3.5}$ are well compatible with normal spiral galaxy progenitors.

\begin{figure}
\centerline{\epsfig{file=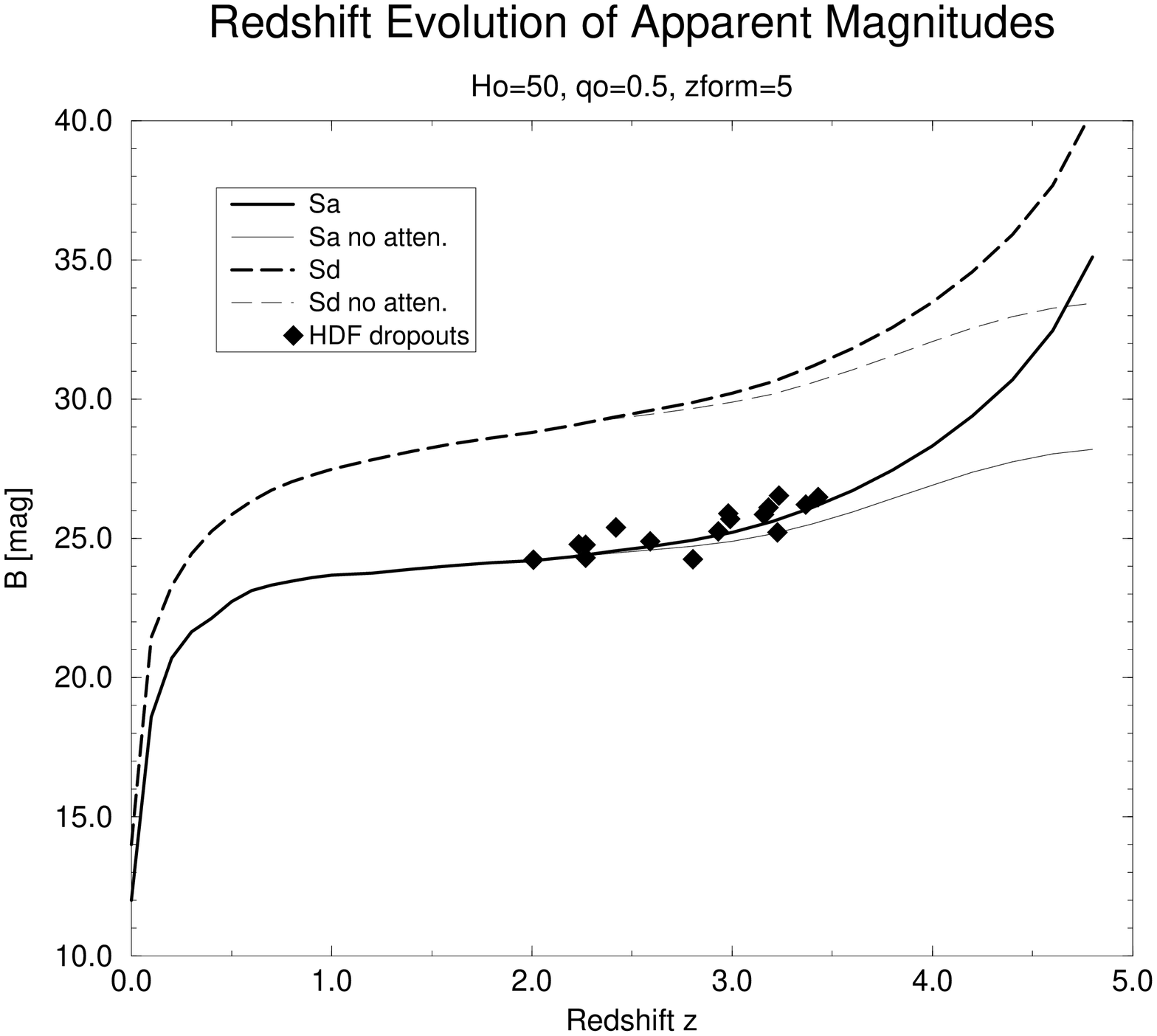,width=12pc,angle=270}\epsfig{file=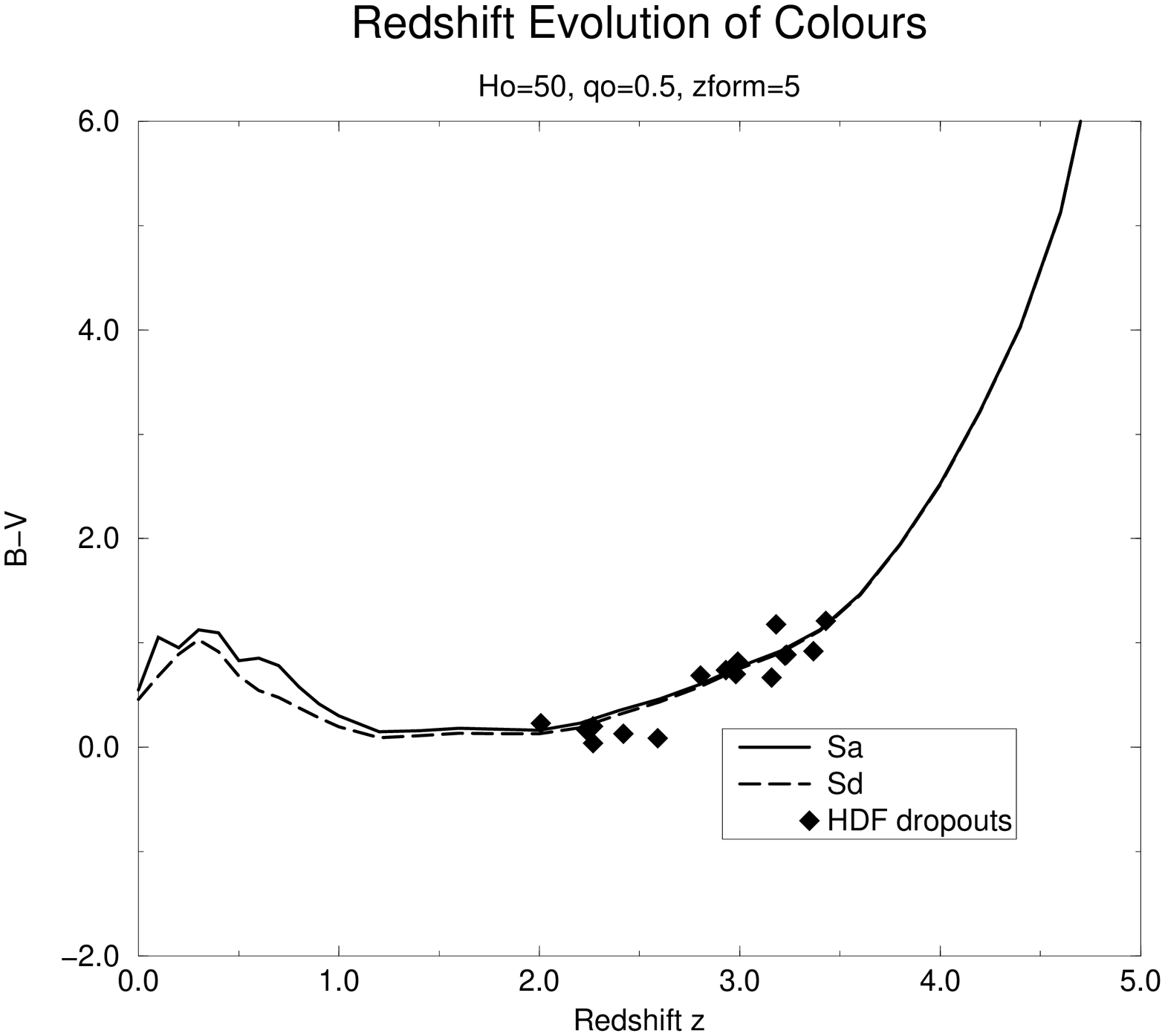,
width=12pc,angle=270}}
\caption{Luminosity and colour evolution of cc spiral models with and without attenuation compared to spectroscopically confirmed LBGs.}
\end{figure}

SFRs derived for LBGs from KECK spectroscopy -- corrected for moderate dust extinction and consistent with their non-detection with SCUBA (Chapman \etal 1999) -- also agree well with spiral model SFRs in this redshift range. These same results are obtained for two different cosmological models ${\rm (H_o=50,~q_o=0.5)}$ and ${\rm (H_o=75,~q_o=0.05)}$, with ${\rm z_f=5}$. The redshift of galaxy formation does not have any visible effect on model luminosities or colours in this redshift range unless 
${\rm z_f < 5}$ for these galaxies. If formed around ${\rm z \sim 5}$, 
classical initial collapse elliptical galaxy models are significantly brighter and have 
SFRs $>20$ times higher than LBGs. 
An extensive comparison including all presently available LBGs is in preparation. 

Clearly, our assumption of one single value for the redshift of galaxy formation is an oversimplification as cosmological structure formation models rather show protracted epochs, extending to very low redshifts for the formation of some galaxy types or masses. No generally accepted specific prediction for the visible parts of galaxies, however, seems to have emerged so far.

\section{SSPs for Various Metallicities}
The time evolution of SSPs of different metallicities, i.e. of stellar systems formed in one short burst of star formation (${\rm t_{SF} = 10^5}$ yr) with one age and one metallicity, is useful not only for the interpretation of star clusters, but also for combination with various kinds of dynamical galaxy evolution or cosmological galaxy formation models, once a SF criterium is specified. Any kind of extended SFH in a galaxy or some part of it -- as complicated as it might be -- is readily expanded into a series of single bursts or SSPs. 

In SSPs (as opposed to galaxies with smooth SFHs), the discreteness of the stellar mass spectrum with evolutionary tracks available causes strong fluctuations in luminosities and colours, that require {\sl a posteriori} smoothing. 
To avoid this without interpolating tracks or using isochrones we developed a Monte Carlo Approach. 

In principle, this method is simple. The HRD population at any timestep is obtained by chosing at random a large number of stellar masses, interpolating their lifetimes, and using for each mass not contained in the original track table appropriate proportions of the two adjacent tracks with each time interval along those two tracks increased or decreased by factors calculated such that the lifetime of the ``artificial star'' equals the interpolated lifetime. 

In Kurth \etal 1999, we present results from our Monte Carlo SSPs. Evolution of luminosities UBVRIJHK, colours, and absorption indices is shown for SSPs of 6 different metallicities ${\rm -2.3 \leq [Fe/H] \leq +0.5}$ over age ranges from $10^7$ yr to 16 Gyr. While early in the evolution of a star cluster, changes in broad band luminosities and colours are generally rapid, they get weaker and weaker with increasing age. SSP models are compared to observations of globular clusters in the Milky Way and M32 and theoretical calibrations for various indices in terms of [Fe/H] are presented in their time evolution. 
Results are available at {\sl http://www.uni-sw.gwdg.de/$\:^{\sim}$ufritze/okurth/ssp.html}. 

Work on the detailed spectral evolution of SSPs from the UV through the NIR is currently in progress. 

\section{Chemically Consistent Chemical Evolution}
Timmes \etal 1995 and Portinari \etal 1998 used stellar yields for a range of 
metallicities to model the chemical evolution of the Milky Way and the solar 
neighbourhood. 
In the following, we will present some of the results from a comparison of our cc chemical evolution models 
with observed abundances in the ISM of {\bf D}amped {\bf L}y$\alpha$ {\bf A}bsorbers ({\bf DLA}s).  

\subsection{CC Spiral Models compared to DLAs}

DLAs show radiation damped Ly$\alpha$ absorption lines in the spectra of background QSOs. These damped Ly$\alpha$ lines are due to high column density gas
(${\rm log~N(HI)~[cm^{-2}] \geq 20.3}$) and usually accompanied by a large number of low ionisation
lines of Al, Si, S, Cr, Mn, Fe, Ni, Zn, ... with the same absorption redshift. 
Only if the often complex velocity structure in the lines can be fully resolved, as e.g. in high-resolution spectroscopy on KECK and WHT with ${\rm \Delta \lambda / \lambda = 60\,000}$, precise element abundances can be derived. 
These are becoming available for a large number of DLAs over the redshift range 0 ... $\geq 4$ (Boiss\'e \etal 1998, 
Lu \etal 1993, 1996, Pettini \etal 1994, 1999, 
Prochaska \& Wolfe 1997, ...). 

Based on similarities between their HI
column densities and those of 
local spiral disks, between their comoving gas densities at high z and densities of 
gas $+$ stars in local galaxies, and based on line
asymmetries indicative of rotation,  DLAs are though to be 
(proto-)galactic disks along the line of sight to
distant QSOs (e.g. Wolfe 1995, Prochaska \& Wolfe 1997, 1998, Wolfe \& Prochaska 1998). Alternatively, Matteucci \etal 1997
propose starbursting dwarf galaxies, Jimenez \etal 1999 LSB galaxies, and Haehnelt \etal 1998
subgalactic fragments to explain DLA galaxies at low and high redshifts, respectively.  

For any cosmological model the time evolution of metallicity Z, element abundances [${\rm X_i/H}$], gas content G, and SFR directly transforms into the corresponding redshift evolution. We stress that for a given SFH and IMF our models yield absolute abundances that do not 
require any scaling or normalisation. 
After referring all observed DLA abundances to one homogeneous set of
oscillator strengths and solar reference values, we compare them with our cc 
chemo-cosmological models in Lindner \etal 1999.

\begin{figure}
\centerline{\epsfig{file=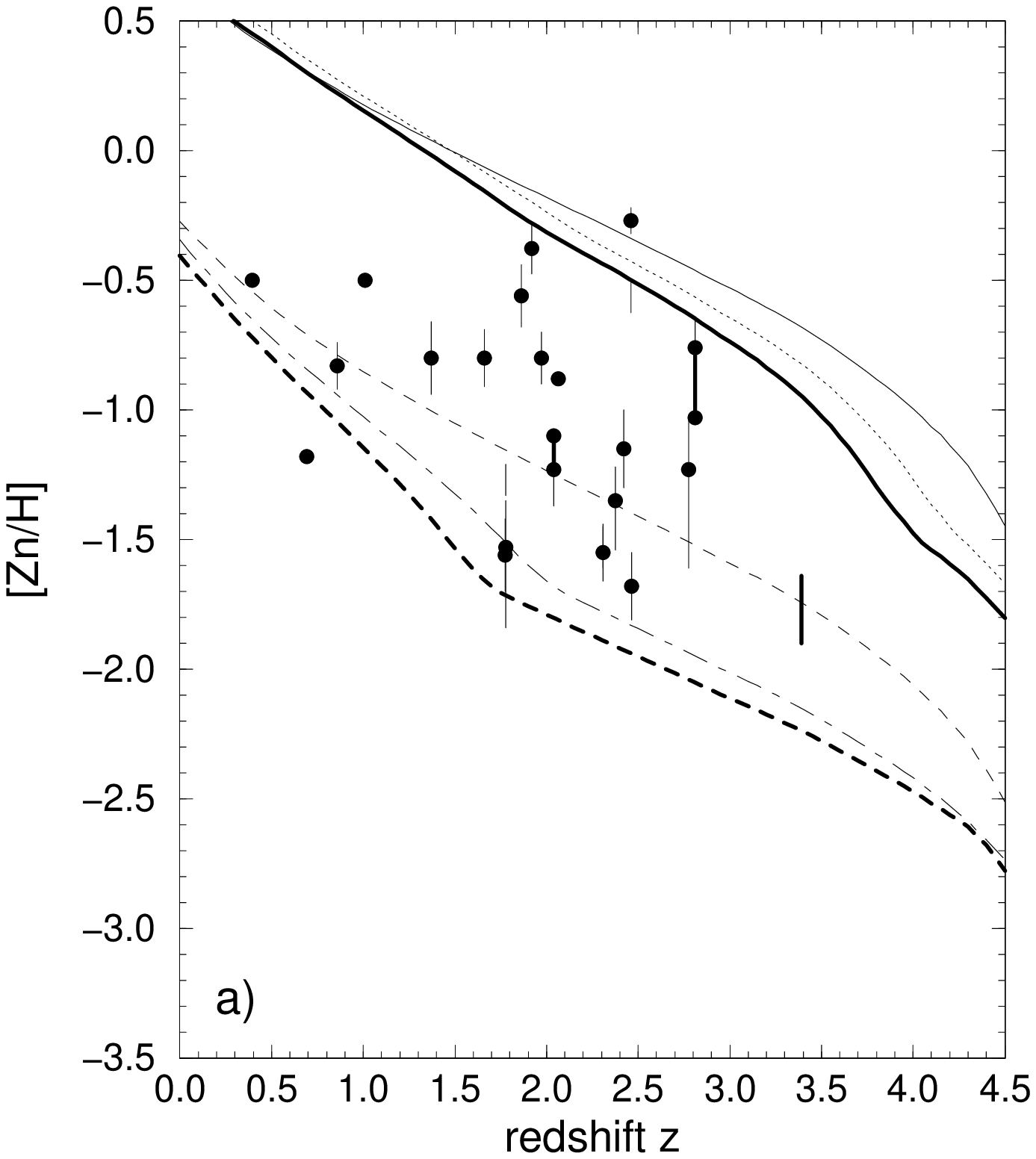,width=15pc}\epsfig{file=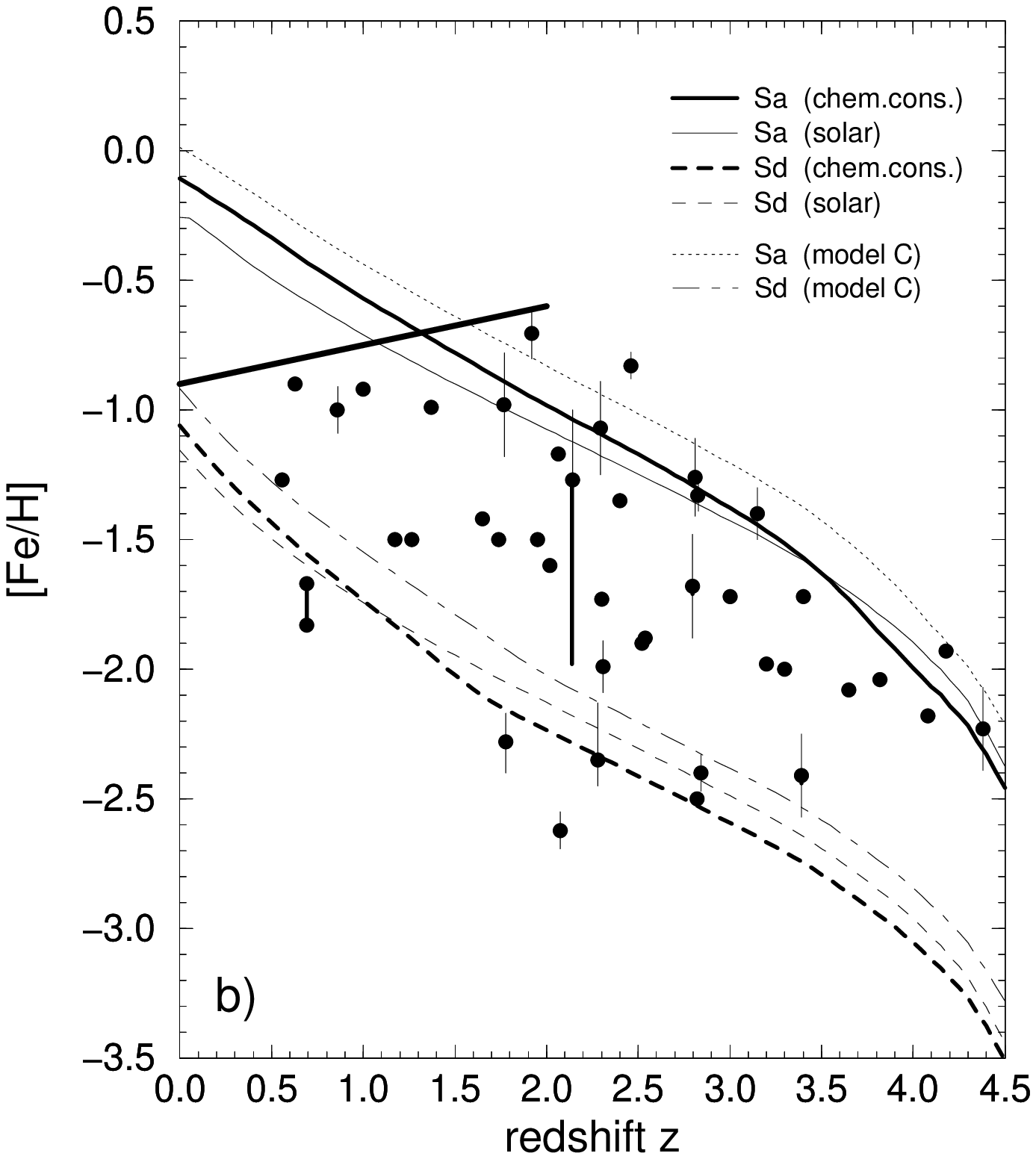,
width=15pc}}
\caption{Redshift evolution of [Zn/H] (2a) and [Fe/H] (2b) for models
 Sa and Sd compared to observed DLA abundances. The straight line in Fig. 2b shows where the model gas content drops below 50\%.}
\end{figure}

For a standard cosmological model (${\rm H_o=50,~\Omega_o=1,~\Lambda_o=0,~z_f=5}$) 
it is seen on the example of Zn, which locally is known not to be depleted on dust grains, 
that our Sa and Sd
models bracket the redshift evolution of DLA abundances from ${\rm z
\geq 4.4}$ to ${\rm z \sim 0.4}$. The entirety of Zn abundances observed in DLAs fall between an upper envelope provided by the rapidly enriching Sa model and a lower envelope made up by the slowly enriching Sd model. The enrichment evolution of Sb and Sc models run between those for Sa and Sd (cf Fig. 2a). 
Similar agreement is found for all
8 elements with a reasonable number of DLA abundances, i.e. for Zn, Fe, Si, Cr, Ni, S, Al, Mn. 
Phillipps \& Edmunds 1996 and Edmunds \& Phillipps 1997 argue that the probability for an arbitrary QSO sightline to cut
through an intervening gas disk and produce DLA absorption is highest
around the effective radius where our models were shown to match average HII abundances in the respective spiral types by ${\rm z = 0}$. Hence, our 
models bridge the gap from high-z DLAs  to 
nearby spirals. {\bf We conclude that from the
point of view of abundance evolution, DLA galaxies may well be the
progenitors of normal spirals Sa -- Sd}, although we cannot exclude 
that some starbursting dwarf or LSB galaxies may also
be among the DLA galaxy sample. If, at the highest redshifts, damped Ly$\alpha$ absorption were caused by subgalactic fragments still bound to merge, our simplified 1-zone models are meant to describe the total SFR of all fragments ending up in one galaxy by low redshift. 
The weak redshift evolution of DLA abundances is a natural result of the 
long SF timescales for disk galaxies. The 
range of SF timescales 
${\rm t_{\ast}}$ for spirals from Sa through Sd fully explains the abundance 
scatter observed among DLAs at any redshift. 

Note that our comparison of spiral galaxy models with DLA abundance data extends to redshifts 
${\rm z \gta 4.4}$, i.e. over lookback times of $> 90 \%$ of the age of the Universe. 

In comparison with solar metallicity models the influence of the metallicity dependent yields is seen to vary from
element to element. Whenever a significant difference is seen, the cc
models give better agreement with the data than ${\rm
Z_{\odot}}$ models. Changing the SN explosion energy (yields for model C from Woosley \& Weaver) has a minor effects (see also Fig. 7a-h in Lindner \etal 1999). 

Somewhat surprisingly, elements which locally deplete strongly onto dust grains (like Fe or Cr) are as well
described by our models as are non-refractory elements like Zn (cf. Fig. 2b). We hesitate, 
however, to draw conclusions about the importance of
dust in DLAs in view of the uncertainties in the stellar yields and the simplicity of our closed-box models. 

We also tried models with constant infall rates and primordial infall abundances. In this case, SFHs have to be adjusted as to still give agreement at ${\rm z=0}$ with average type-specific HII region abundances, colours, etc. This is the reason for moderate constant infall rates to only marginally change the redshift evolution of model abundances and to not affect any of our conclusions. A redshift dependent infall rate, however, with or without evolving infall abundance, might affect our results. The difficulty is to constrain the additional parameter(s) without embedding the galaxy into a cosmological context. Mass estimates from column densities, linear dimensions (DLAs seen in close QSO pairs), and rotation velocities (Prochaska \& Wolfe 1997) indicate that -- at least some of -- these systems at ${\rm z \sim 2 - 3}$ already have the masses of local spirals and, hence, may not require important infall. 

\subsection{A Change with Redshift in the DLA Population ?}

While at high redshift all spiral types Sa through Sd seem to
give rise to DLA absorption, no more data points at ${\rm z \lta 1.5}$
reach close to our early type spiral models. The boundary to the observed abundances coincides with the 50 \% gas-to-total-mass ratio in our models. I.e., by low redshift, the gas poor
early type spirals seem to drop out of DLA samples. A deficiency of high N(HI) DLA systems 
at low z had been noted before (Lanzetta \etal 1997) and attributed to their high metallicity 
and dust content (Steidel \etal 1997). We give an additional reason: as the global 
gas content drops, the probability for an arbitrary line of sight to a QSO to cut through a high column density part of the 
galaxy decreases, i.e. the cross section for {\bf damped} Ly$\alpha$ absorption gets reduced. 

\subsection{Spectroscopic Predictions for DLA Galaxies \\
and Future Prospects}
Despite considerable efforts and large amounts of telescope time devoted to the optical identification of galaxies responsible for damped Ly$\alpha$ absorption, only less than a handful have been found by today. Contrary to earlier expectations, the success rate has not increased when low redshift DLAs were found in UV spectra of QSOs. If, indeed, there were a shift in the DLA galaxy population towards later spiral types at lower redshift, then this is what our models predict since, locally on average, Sd galaxies are fainter by 
$\sim 2$ mag in B than Sa's. Hence, the low-z late type DLA 
galaxies should be about as faint in B, \R, and K as the brightest members (i.e. the early spiral types) of the high-z population. Both, an average Sd model at ${\rm z \sim 0.5}$ and an average Sa model at ${\rm z \sim 2 ~-~ 3}$ have ${\rm B \sim 25}$, \R $\sim 24.5$, K $\sim 22$ mag. Luminosities of
the few optically identified DLA galaxies (and candidates) are
in good agreement with our predictions. 
In particular, if there were early type spirals among the low redshift DLA galaxy population they would have ${\rm B \sim 22.5}$, \R $\sim 21$, K $\sim 18.5$ mag at ${\rm z \sim 0.5}$ and they would have been detected. Their non-detection is consistent with our finding of a change with redshift in the DLA galaxy population (see F.-v.A. \etal 1999a, b, c for details). 

DLA galaxies are within the reach of 10m-class 
telescopes up to redshifts z $> 3$ and trace the normal galaxy population to these high 
redshifts without any bias as to high luminosity, radio power, or the like. With information 
about ISM abundances from the metal absorption lines {\bf and}
spectrophotometric properties of the stellar population, they will allow for powerful 
constraints on model parameters, ages, and cosmological parameters (cf. Lindner \etal 1996). 
Accurate abundance data in very low metallicity DLAs may 
provide clues for the nucleosynthesis at low metallicity. 

The knowledge about the abundance evolution of gaseous spiral disks as a function of redshift will allow to predict abundances of stars, star clusters, and Tidal Dwarf Galaxies that form in the powerful starbursts accompanying spiral galaxy mergers in the local universe and in the past (cf. Sect. 6.3 and 6.4). As will be shown, knowing these abundances is important for properly interpreting observed colours and luminosities of star clusters and Tidal Dwarf Galaxies in terms of ages, masses, etc. Spectroscopy of those objects, in turn, provides an independent cross check of our abundance predictions.

\section{Interactions and Starbursts}
\subsection{Importance of Galaxy Interactions and Starbursts}
Evidence for the importance of galaxy interactions and merging is coming both from theoretical galaxy formation modelling and observations. 

In a variety of cosmological contexts (CDM with and without $\Lambda$, mixed DM), semianalytic as well as numerical structure formation models predict galaxies to build up hierarchically from a series of mergers  of successively larger masses (see e.g. Carlberg 1990, Lacey \& Cole 1993, Kauffmann \& White 1993). With increasing redshift, observations indicate increasing numbers of physically close galaxy pairs (e.g. Zepf \& Koo 1989), galaxies seen in interaction, and disturbed galaxy morphologies indicative of a recent interaction (e.g. Conselice \& Bershady 1999). All strong galaxy interactions or mergers are observed to be accompanied by bursts of star formation if one or both of the galaxies are gas rich. 
When falling into a cluster, the spiral rich field galaxy population must somehow be transformed into the E/S0/dE rich cluster population. These transformation processes also involve interactions -- among individual galaxies or of an infalling galaxy with the cluster environment, 
potential, or its hot intracluster medium. In gas rich galaxies some of these processes, as well, may trigger starbursts -- sometimes powerful and global, as reported for E$+$A galaxies by Poggianti \& Barbaro 1996. 

In dissipationless stellardynamical mergers (of disks or spheroidal systems), the remnant after partially complete violent relaxation usually is a spherical system. Its central phase space density cannot significantly increase beyond that of the progenitors. Therefore, it does not seem possible to produce the high central density cores of massive ellipticals  by dissipationless mergers of less massive subcomponents or dwarf galaxies which intrinsically have lower central densities. 
Gaseous mergers, however, are highly dissipative, the gas is driven to the center very efficiently, and gas densities comparable to the central stellar densities in massive ellipticals are reached in high-resolution simulations. In Ultraluminous Infrared Galaxies ({\bf ULIRGs}), all of which have been shown to be mergers of gas rich spirals, central molecular gas densities of the order of 1000 ${\rm M_{\odot} \, pc^{-3}}$ are indeed observed. 

It has not been possible yet to treat the full dynamics of mergers involving stars, DM, and gas in at least 3 phases (Hot X gas, HI, cold molecular gas), and to consistently include SF and feedback (under extreme conditions) into high-resolution stellar $+$ gaseous dynamical models of interacting gas rich galaxies. It is clear, however, that strong bursts of SF or/and efficient fuelling of a central AGN may occur. Which alternative is chosen or dominates may depend on the properties of the galaxies involved, the geometry of the encounter, and possibly on the stage of the interaction (see e.g. F.-v.A. 1996 for a review in Galaxy Interactions). 

By studying interaction-triggered starbursts with spectrophotometric and chemical evolutionary synthesis models we hope to learn about the SF process, SF efficiencies, etc. under the violent conditions in mergers/strong interactions that seem to significantly differ from the comparatively ``peaceful'' SF situation in our Milky Way.

\subsection{Selected Results}
We calculated an extensive grid of starburst models. In the idealisation of equal-type mergers Sa-Sa, Sb-Sb, Sc-Sc, Sd-Sd, occuring at a variety of evolutionary stages of the parent galaxies, we model starbursts of various strengths and durations on top of the pre-existing stellar population. 
The age and type of the progenitor galaxies determine the luminosities and colours of the stellar population and the ISM metallicity at the onset of the burst. 
The gas content of the spirals at merging sets an upper limit to the gas reservoir available for SF in the burst. Burst strengths being defined as the increase of the stellar mass S during the burst, ${\rm b := \frac{\Delta S_{burst}}{S_{pre-burst}}}$, this means that 
e.g. Sa-Sa mergers in the local universe cannot have bursts as strong as Sd-Sd mergers today or Sa-Sa mergers earlier in their evolution when the galaxies contained more gas. 

The evolution of galaxies before, during, and after the burst was studied in terms of two-colour diagrams, as well as in their luminosity and metallicity evolution for solar metallicity models in F.-v.A. \& Gerhard 1994a. 

The models show that after a burst spiral-spiral mergers of galaxy types Sa, Sb, Sc occuring 3 -- 4 Gyr ago readily develop typical elliptical galaxy colours by today -- provided SF ceases completely. Sd mergers remain too blue for $> 4$ Gyr. In detail, the timescale until a merger remnant with its postburst reaches the observed colour range of E/S0s depends on the progenitor types, the burst strength, and on the wavelength range considered. 

Using our grid of models we analysed the starburst in the prototypical merger remnent NGC 7252, for which a wealth of additional information is available (e.g. spectra from UV -- optical, HI-, HII-, CO-maps, dynamical modelling). 
Despite its enormous tidal tails it already features 
an azimutally averaged ${\rm r^{1/4}}$ -- profile in its inner part, and it is the oldest from Toomre's (1977) dynamical age sequence of interacting galaxies. 
Comparing broad band colours UBVR with our model grid, we were left with some range of possible combinations of burst strengths and ages. Within this cell of parameter space, we in detail compared our model spectra with the spectrum of NGC 7252 showing deep Balmer absorption lines and a small H$_{\beta}$ emission component. This 
allowed us to identify a very strong starburst that increased the stellar population by $\lta 50$ \% and started $\sim 1$ Gyr ago, i.e. around the time of ${\rm 1^{st}}$ pericenter as obtained from dynamical modelling (cf. Hibbard \& Mihos 1995). Both in terms of morphology and colours, NGC 7252 could evolve into an elliptical galaxy within the next 1 -- 3 Gyr, provided its SFR $\rightarrow 0$. 
Its present centrally concentrated SFR as obtained from an IUE spectrum (F.-v.A. \etal {\sl in prep.}) is ${\rm \sim 3~ M_{\odot}yr^{-1}}$, consistent with the weak H$_{\beta}$ emission. HI falling back from the tidal tails (Hibbard \& Mihos 1995) together with gas restored by dying burst stars supports the present SFR and may do so for another few Gyrs, allowing NGC 7252 to rebuild itself a small disk of stars and gas. In this case, both in terms of morphology and colours, it might rather come to resemble an S0 or even an Sa galaxy. 

In an attempt to identify typical values and ranges for burst parameters and study their possible dependence on galaxy and interaction properties, Liu \& Kennicutt's (1995) sample of interacting galaxies was analysed by O. Kurth in his diploma thesis (G\"ottingen 1996). Only on the basis of those typical values and relations can interaction induced starbursts consistently be included into the statistical evolution of high redshift galaxies, as e.g. into models for the redshift evolution of the luminosity function or into cosmological galaxy formation scenarios. 
In general, burst durations can only weakly be constrained to be of the order of a few $10^8$ yr for spiral -- spiral mergers, i.e. definitely longer than for Blue Compact Dwarf 
Galaxies ({\bf BCDs}), where they are of the order of $10^6$ yr. In both cases, burst durations reflect the dynamical timescales of the systems involved. 
One of the most important results is that {\bf starbursts in galaxy mergers can be very strong}, increasing the stellar mass by as much as 30 \% or even more. In all massive interacting systems, burst strengths $\gta 10 \%$ were found, larger by a factor $10 - 100$ than the burst strengths obtained for BCDs (Kr\"uger \etal 1995). 

For NGC 7252, the peak SFR during the burst must have been several hundred ${\rm M_{\odot}\,yr^{-1}}$, comparable to those of ULIRGs. Assuming a gas content from the high end of what is typical for Sc galaxies, the SF efficiency ({\bf SFE}) -- defined as the mass of stars formed in the burst relative to the mass of gas available ${\rm \eta := \frac{\Delta S_{burst}}{G_{pre-burst}}}$ -- must have been very high, of order 30 -- 45 \%, a factor $\gg 10$ larger than in the Milky Way (F.-v.A. \& Gerhard 1994b). 

CO, HCN, and CS observations of ULIRGs tracing molecular gas at densities ${\rm n \sim 500,~10^4,~10^5~cm^{-3}}$, respectively, explain their high SFEs as a consequence of their high fractions of molecular gas at the highest densities (Solomon \etal 1992). 
While for BCDs a trend for burst strengths to decrease with increasing galaxy mass was found (Kr\"uger \etal 1995), burst strengths and SFEs in interacting galaxies with masses 1 -- 2 orders of magnitude higher are generally higher by the same 1 -- 2 orders of magnitude. Does this mean, that the SF process is intrinsically different in different environments ? (See Sect. 5 in F.-v.A. 1996 for a discussion of ``violent vs peaceful'' SF and the possible relation with differences in the molecular cloud structure). 

SFEs that high are usually assumed for the collapse of protogalactic clouds in the early Universe that gives rise to globular cluster formation. We speculated that 
globular cluster formation might have been possible in the starburst of NGC 7252 and predicted the metallicity of both stars and clusters forming in the burst to be ${\rm Z_{\ast} \geq Z_{ISM}^{Sc(t_o - 1~Gyr)} \gta \frac{1}{2} \cdot Z_{\odot}}$. 

\subsection{Young Star Clusters in Mergers}
Within a few years, bright {\bf Y}oung {\bf S}tar {\bf C}luster ({\bf YSC}) populations were detected with HST in NGC 7252 and a number of other interacting galaxies and merger remnants. They are interesting for many reasons. E.g., star clusters are better suited than the integrated light to study the age and duration of the starburst. Being most probably observed around the same galactocentric radii at which they formed, they thus allow to study the spatial extent of the starburst and -- for sufficiently large cluster populations -- its time evolution. YSCs offer the possibility to study star and cluster formation processes in a violent environment, in particular in comparison with observations of the molecular gas content or cloud structure. 
Using SSP models for the interpretation of YSC systems in the old merger remnant NGC 7252 (F.-v.A. \& Burkert 1995) and in the young interacting Antennae galaxies NGC 4038/39 (F.-v.A. 1998a), we show that metallicities of YSCs are crucial to age-date them on the basis of observed broad band colours and to predict their future luminosity and colour evolution. For a few of the brightest YSCs in NGC 7252, our metallicity prediction from spiral models was confirmed by spectroscopy (Schweizer \& Seitzer 1993, 1998). 

The question if some, many, or most of these YSCs could be young {\bf G}lobular {\bf C}lusters ({\bf GC}s) is of eminent importance. If a secondary GC population, comparable in number to the original one, can be formed in a merger induced starburst, this would invalidate the last surviving argument against the formation of -- at least some -- elliptical galaxies from spiral-spiral mergers based on the difference in specific GC frequencies between elliptical galaxies and spirals. 
Analysing WFPC1 data for some 40 YSCs in NGC 7252 and $> 500$ in NGC 4038/39 we tentatively conclude that the bulk of the YSC populations in both systems may well evolve into ``normal'' GC populations in terms of colour distributions, luminosity and mass functions (F.-v.A. 1999a). Before definite conclusions can be reached, we are currently reanalysing WFPC2 data and studying YSCs in an age sequence of interacting, merging and merged galaxies to assess the impact of dynamical cluster destruction processes. 

Because of its enhanced metallicity, a secondary GC population might eternally testify to a merger origin, still at times when tidal tails, kinematic peculiarities, and fine structure  will long have disappeared, and colours will no longer reveal a past starburst. Analyses of GC systems in a number of elliptical galaxies -- from low-luminosity ellipticals all through cD galaxies -- have revealed bimodal colour distributions in many bright ellipticals, including 2 S0s, broad or multi-peak colour distributions in all cD galaxies investigated, and single-peak distributions in a few low-luminosity ellipticals (Kissler -- Patig {\sl this volume}, Kissler -- Patig \etal 1998, Gebhardt \& Kissler -- Patig 1999). 
With 10 m telescopes in combination with HST imaging, MOS becomes feasible for YSCs and even old GCs out to Virgo distances. In comparison with SSP models, it will allow to decompose the colour distribution of star cluster systems into metallicity and age distributions and, thereby, give information about the formation of their parent galaxies. 

\subsection{Tidal Dwarf Galaxies} 
Not only secondary populations of YSCs are formed during mergers of gas-rich galaxies, but also a new galaxy formation mechanism has been detected in these systems a few years ago. 
In the extended tidal tails of several interacting systems bright blue star forming knots, often associated with large HI concentrations, are observed with masses and luminosities typical of dwarf galaxies. Two of these {\bf T}idal {\bf D}warf {\bf G}alaxies ({\bf TDG}s) are detected in NGC 7252, one in NGC 4038/39, two in Arp 105, several in the ``Superantennae'', and other systems (see Duc \etal 1998 for a recent review). 

TDGs form from ``recycled'' material torn out from a spiral disk. They deviate from the dwarf galaxy luminosity -- metallicity relation in the sense that they have enhanced metallicities for their luminosities as compared to ``non-recycled'' dwarf galaxies. Abundance determinations from their HII region-like spectra agree well with predictions from our spiral models. 

While TDGs forming in the lower parts of tidal features are expected to fall back into the merger remnant on timescales of $10^8$ to few $10^9$ yr, those at the tips of the tails will probably escape and might survive as independent entities. Kinematic independence from the parent galaxy has been demonstrated in a few cases (Duc 
\etal 1997, Duc \& Mirabel 1998). 

Independently, both N-body simulations for the stars and hydrodynamical models for the gas in merging galaxies feature condensations along their respective tidal tails, leading to two competing scenarios for the formation of TDGs (cf. Barnes \& Hernquist 1992, Elmegreen \etal 1993).

Evolutionary synthesis models involving fractions of the stellar populations of spiral progenitors plus starbursts of various strengths are used to analyse the stellar populations and the evolutionary states of TDGs. Gaseous continuum and line emission are included. 
Analysing a small sample of TDGs observed by P.-A. Duc with a coarse grid of models (F.-v.A. \etal 1998) we find evidence for strong bursts, ${\rm b := \frac{S^{young}}{S^{old}} = 0.1 - 0.4}$, on top of `old' stellar populations, i.e. populations with the age distribution of stars in their parent spiral (${\rm S^{young}~and~S^{old}}$ being the masses of young and 'old' stars, respectively). Since these strong bursts completely dominate the light in the optical, NIR data are required to constrain the mass contribution of the `old' population. The latter is important for the fading that TDGs will experience in the future, as well as for their dynamical evolution. Without a significant old population the strongest bursts might disrupt the TDG. Both, the fading and future dynamical evolution, in turn, are relevant for a possible cosmological significance of TDGs. In the past, the merger rate was higher, galaxies were more gas-rich and probably less stable so that the production of a significant population of TDGs might be expected. 

To explain that part of the faint blue galaxy excess that is due to dwarf galaxies, Babul \& Ferguson 1996 invoke a hypothetic population of dwarf galaxies, the formation of which is delayed until ${\rm z \sim 1}$ by the intergalactic radiation field. Properties they require for their dwarf galaxies to explain the faint blue galaxy excess are very similar to those of our TDGs. Depending on their fading, TDGs might well explain part of the faint blue galaxy excess. Moreover, the remnant problem faced by Babul \& Ferguson and Ferguson \& Babul 1998 would be alleviated if part of the TDGs would spiral back into the merger remnant on timescales of a $10^8 - 10^9$ yr.

In Weilbacher \etal 2000 ({\sl submitted}) we analyse a sample of 10 interacting galaxies from the AM catalogue with imaging in B, V, R. Comparing colours of knots along tidal structures with an extensive 
grid of TDG model calculations, a number of promising TDG candidates are identified. Knots with colours not explained by models most probably are background galaxies. TDG candidates in this first sample have young burst ages of $\sim 7$ Myr, burst strengths ${\rm b \lta 0.2}$, and will fade by up to 2.5 mag in B, on average, within 200 Myr after the burst. 
Follow-up spectroscopy with VLT and HET is underway, models are currently being calculated for their detailed spectral evolution. 
In any case, this recently discovered mode of ongoing galaxy formation in the local Universe from recycled gas and stars is an interesting field of study just at the beginning of being explored.

\section{Conclusion and Outlook}
I presented a versatile model for the evolution of stellar populations and gas that offers a variety of applications from star clusters to nearby and distant galaxies. Only a few of them have been presented here. 
With a minimum number of parameters our combined chemical, spectrophotometric and cosmological evolution models describe a large number of observables and provide a long evolutionary baseline to compare with high-redshift galaxy abundances and spectra and understand the evolution of various galaxy types from the earliest phases to the present. 
The chemically consistent treatment is a first attempt to consistently combine 2 out of 3 aspects of galaxy evolution that nature, too, has coupled. 

Interactions play an important role for the evolution of galaxies over cosmological timescales. While in its present state, the model does not include any dynamical aspect nor spatial resolution, we tried and studied the effects of starbursts accompanying galaxy interactions if gas is involved. Several interesting phenomena were observed in this context, as e.g. the formation of large populations of bright star clusters and of ``recycled'' Tidal Dwarf Galaxies. Application of our models provided a first step to understand the nature and properties of these systems as well as their possible future evolution. 

The model has allowed for a series of precise observational predictions, part of which became verified already while others keep standing for a test. 

Our first attempt to also include the ${\rm 3^{rd}}$ aspect of galaxy evolution, the formation and dynamical evolution of galaxies in their cosmological environments, could not be discussed here (cf. Contardo \etal 1998).  

Over the years this model was developed, extended and refined, with its input physics continuously updated, observational extragalactic research has seen a tremendous progress. The amount of information
from HST and large ground-based surveys is enormous and several quantum steps have been performed, e.g. concerning image resolution with HST, spectral resolution with KECK and WHT, and the number of high-redshift galaxies by the Lyman break technique. With 10 m telescopes, like VLT and HET, observational progress is challenging theory to keep path. 
The particularly close interplay between observations and the conceptually simple galaxy evolution modelling presented here, has proven very fruitful and stimulating for both sides.

\medskip\noindent
{\bf Acknowledgements} {\footnotesize

\noindent
My thanks go to K. Fricke for his encouragement and to all the present and former members of our Galaxy Evolution Group, who -- over the years -- have contributed to various aspects of the work presented here, i.e. to Harald Kr\"uger, 
Christian Einsel, Johannes Loxen, Claudia M\"oller, Oliver Kurth, Ulrich Lindner, Peter Weilbacher, Jens 
Bicker, and Jochen Schulz. 
Partial financial support from the {\sl Deutsche For\-schungs\-gemeinschaft} and the {\sl Verbundforschung Astronomie} for various aspects of this work is gratefully acknowledged. }

\vspace{0.7cm}
\noindent

\vfill


{\footnotesize

\begin{thebibliography}{98}

\bibitem{} Babul, A., Ferguson, H. C., 1996, ApJ 458, 100

\bibitem{}  Barnes, J. E., Hernquist, L., 1992, Nat. 360, 715

\bibitem{}  Boiss\'e, P., Le Brun, V., Bergeron, J., Deharveng, J.-M., 1998, A\&A 333, 841

\bibitem{} Bressan, A., Fagotto, F., Bertelli, G., Chiosi, C., 1993, A\&AS 100, 647

\bibitem{}  Bruzual, G. A., Charlot, S., 1993, ApJ 405, 538

\bibitem{}  Carlberg, R., 1990, ApJ 350, 505

\bibitem{}  Chabrier, G., Barafffe, I., 1997, A\&A 327, 1039

\bibitem{} Chapman, S. C., Scott, D., Steidel, C. C., \etal 1999, ({\sl astro-ph/9909092})

\bibitem{}  Conselice, C. J., Bershady, M., 1999, ApJ {\sl in press} ({\sl astro-ph/9907399})

\bibitem{}  Contardo, G., Steinmetz, M., Fritze -- v. Alvensleben, U., 1998, ApJ 507, 497

\bibitem{}  Duc, P.-A., Mirabel, I. F., 1998, A\&A 333, 813

\bibitem{}  Duc, P.-A., Mirabel, I. F., Brinks, E., 1998, Highlights of Astron. 11B, 141

\bibitem{}  Duc, P.-A., Brinks, E., Wink, J. E., Mirabel, I. F., 1997, A\&A 326, 537

\bibitem{}  Edmunds, M. G., Phillipps, S., 1997, MN 292, 733

\bibitem{}  Elmegreen, B., Kaufmann, M., Thomasson, M., 1993, ApJ 412, 90

\bibitem{} Fagotto, F., Bressan, A., Bertelli, G., Chiosi, C., 1994, A\&AS 104, 365, A\&AS 105, 29$+$39

\bibitem{}  Ferguson, A. M. N., Gallagher, J. S., Wyse, R. F. G., 1998, AJ 116, 673

\bibitem{}  Ferguson, H. C., Babul, A., 1998, MN 296, 585

\bibitem{}  Fria\c ca, A. C. S., Terlevich, R. J., 1999, MN 305, 90

\bibitem{}  Fritze - v. Alvensleben, U., 1989, PhD Thesis, Univ. G\"ottingen

\bibitem{}  Fritze - v. Alvensleben, U., 1994, in {\sl Panchromatic View of Galaxies: Their Evolutionary Puzzle}, ed. G. Hensler, Editions Fronti\`eres, p. 245

\bibitem{}  Fritze - v. Alvensleben, U., 1996, in {\sl From Stars to Galaxies: the Impact of Stellar Physics on Galaxy Evolution}, eds. C. Leitherer, U. Fritze -- v. Alvensleben, J. Huchra, ASP Conf. Ser. 98, p. 496

\bibitem{}  Fritze - v. Alvensleben, U., 1998a, A\&A 336, 83

\bibitem{}  Fritze - v. Alvensleben, U., 1998b, Highlights of Astron. 11A, 78

\bibitem{}  Fritze - v. Alvensleben, U., 1999a, A\&A 342, L25

\bibitem{}  Fritze - v. Alvensleben, U., 1999b, in {\sl Age Dating of Stars and Galaxies}, 
eds. I. Hubeny, S. Heap, B.  Cornett, ASP Conf. Ser. 192, 273

\bibitem{}  Fritze - v. Alvensleben, U., Burkert, A., 1995, A\&A 300, 58

\bibitem{}  Fritze - v. Alvensleben, U., Gerhard, O.E., 1994a, A\&A 285, 751

\bibitem{}  Fritze - v. Alvensleben, U., Gerhard, O.E., 1994b, A\&A 285, 775

\bibitem{}  Fritze - v. Alvensleben, U., M\"oller, C. S., Duc, P.-A., 1998, in {\sl Dwarf Galaxies and Cosmology}, eds. T. X. Thuan, C. Balkowski, J. Tran Thanh Van, 
Editions Fronti\`eres, {\sl in press}, (astro-ph/9805330)

\bibitem{}  Fritze - v. Alvensleben, U., Lindner, U., M\"oller, C. S., 1999a, in {\sl Chemical
Evolution from Zero to High Redshift}, ed. J. Walsh, M. Rosa, Springer Berlin, p. 256

\bibitem{}  Fritze - v. Alvensleben, U., Lindner, U., M\"oller,
C. S., Fricke, K. J., 1999b, in {\sl The Evolution of Galaxies on
Cosmological Timescales}, ed. J. Beckman, {\sl in press}, (astro-ph/9904208)

\bibitem{}  Fritze - v. Alvensleben, U., Lindner, U., M\"oller,
C. S., 1999c, in {\sl Building Galaxies: From the Primordial Universe
to the Present}, eds F. Hammer \etal, Editions Fronti\`eres, {\sl in press}, (astro-ph/9907396)

\bibitem{}  Gebhardt, K., Kissler -- Patig, M., 1999, AJ 118, 1526

\bibitem{}  Giavalisco, M., Steidel, C. C., Macchetto, F. D., 1996, ApJ 470, 189

\bibitem{}  Haehnelt, M. G., Steinmetz, M., Rauch, M., 1998, ApJ 495, 647

\bibitem{}  Hibbard, J. E., Mihos, J. C., 1995, AJ 110, 140

\bibitem{}  Hibbard, J. E., Vacca, W. D., 1997, AJ 114, 1741

\bibitem{}  v. d. Hoek, L. B., Groenewegen, M. A. T., 1997, A\&AS 123, 305

\bibitem{}  Jimenez, R., Bowen, D.V., Matteucci, F., 1999, ApJ 514, L83

\bibitem{}  Kauffmann, G., White, S. D. M., 1993, MN 261, 921

\bibitem{}  Kennicutt, R. C., 1992, ApJS 79, 255

\bibitem{} Kinney, A. L., Calzetti, D., Bohlin, R. C., \etal, 1996, ApJ 467, 38

\bibitem{}  Kissler -- Patig, M., Forbes, D. A., Minniti, D., 1998, MN 298, 1123

\bibitem{}  Kobayashi, C., Tsujimoto, T., Nomoto, K., Hachisu, I.,
Kato, \etal, 1998, ApJ 503, L155

\bibitem{}  Kr\"uger, H., Fritze - v. Alvensleben, U., Loose, H.-H., 1995, A\&A 303, 41

\bibitem{}  Kurth, O., Fritze - v. Alvensleben, U., Fricke, K. J.,
1999, A\&AS 138, 19

\bibitem{}  Lacey, C., Cole, S., 1993, MN 262, 627

\bibitem{}  Lanzetta, K. M., Wolfe, A. M., Altan, H., \etal, 1997,
AJ 114, 1337

\bibitem{}  Lejeune, T.,  Cuisinier, F., Buser, R., 1997, A\&AS 125, 229

\bibitem{}  Lejeune, T.,  Cuisinier, F., Buser, R., 1998, A\&AS 130, 65

\bibitem{}  Lindner, U., Fritze - v. Alvensleben, U., Fricke, K. J.,
1996, A\&A 316, 123

\bibitem{}  Lindner, U., Fritze - v. Alvensleben, U., Fricke, K. J., 1999, A\&A 341, 709

\bibitem{}  Liu, C. T., Kennicutt, R. C., 1995, ApJS 100, 325

\bibitem{}  Lowenthal, J. D., Koo, D. C., Guzman, R., \etal, 1997, ApJ
481, 673

\bibitem{}  Lu, L., Wolfe, A. M., Turnshek, D. A., Lanzetta, K. M., 1993, ApJS 84, 1

\bibitem{}  Lu, L., Sargent, W. L. W., Barlow, T. A., \etal, 1996, ApJS 107, 475

\bibitem{}  Madau, P., 1995, ApJ 441, 18

\bibitem{}  Madau, P., Ferguson, H. C., Dickinson, M. E., \etal, 1996, MN 283, 1388

\bibitem{}  Matteuci, F., Greggio, L., 1986, A\&A 154, 279

\bibitem{}  Matteucci, F., Molaro, P., Vladilo, G., 1997, A\&A 321, 45

\bibitem{}  McWilliam, A., Rich, R. M., 1994, ApJS 91, 749 

\bibitem{}  M\"oller, C. S., Fritze - v. Alvensleben, U., Fricke,
K. J., 1997, A\&A 317, 676

\bibitem{}  M\"oller, C. S., Fritze - v. Alvensleben, U., Fricke,
K. J., 1998,  in {\sl The Birth of Galaxies}, {\sl in press} 

\bibitem{}  Nomoto, K., Iwamoto, K., Nakasto, N., \etal, 1997,
Nucl. Phys. A, A621

\bibitem{}  Oey, M. S., Kennicutt, R. C., 1993, ApJ 411, 137

\bibitem{}  Pettini, M., Smith, L. J., Hunstead, R. W., King, D. L.,
1994, ApJ 426, 79

\bibitem{}  Pettini, M., Smith, L. J., King, D. L., Hunstead, R. W., 
1997, ApJ 486, 665

\bibitem{}   Pettini, M., Ellison, S. L., Steidel, C. C., Bowen,
D. V., 1999, ApJ 510, 576

\bibitem{}  Phillipps, S., Edmunds, M. G., 1996, MN 281, 362

\bibitem{} Poggianti. B. M., Barbaro, G., 1996, A\&A 314, 379

\bibitem{}  Portinari, L., Chiosi, C., Bressan, A., 1998, A\&A 334, 505

\bibitem{}  Prochaska, J. X., Wolfe, A. M., 1997, ApJ 487, 73

\bibitem{}  Prochaska, J. X., Wolfe, A. M., 1998, ApJ 507, 113

\bibitem{}  Richer, M. G., McCall, M., 1995, ApJ 445, 642 

\bibitem{}  Rocca -- Volmerange, B., Guiderdoni, B., 1988, A\&AS 75, 93

\bibitem{}  Rocha - Pinto, H. J., Maciel, W. J., 1998, A\&A 339, 791

\bibitem{}  Sandage, A., Binggeli, B., Tammann, G. A., 1985, AJ 90, 395 $+$ 1795

\bibitem{}  Scalo, J. M., 1986, Fundam. Cosmic Phys. 11, 1

\bibitem{}  Schweizer, F., Seitzer, P., 1993, ApJ 417, L29

\bibitem{}  Schweizer, F., Seitzer, P., 1998, AJ 116, 2206

\bibitem{} Solomon, P. M., Downes, D., Radford, S. J. E., 1992, ApJ 387, L55

\bibitem{}  Somerville, R. S., Primack, J. R., Faber, S. M., 1998, {\sl astro-ph/9806228}

\bibitem{}  Steidel, C. C., Pettini, M., Hamilton, D., 1995, AJ 110, 2519

\bibitem{}  Steidel, C. C., Giavalisco, M., Pettini, M., \etal, 1996, ApJ 462, L17

\bibitem{}  Steidel, C. C., Dickinson, M., Meyer, D. M., \etal, 1997, ApJ 480, 568

\bibitem{}  Steidel, C. C., Adelberger, K. L., Giavalisco, M., Dickinson, M., 1999, ApJ 519, 1

\bibitem{}  Timmes, F. X., Woosley, S. E., Weaver, T. A., 1995, ApJS 98, 617

\bibitem{}  Tinsley, B. M., 1980, Fundam. Cosmic Phys. 5, 287

\bibitem{}  Toomre, A., 1977, in {\sl The Evolution of Galaxies and Stellar Populations}, eds. B. Tinsley, R. B. Larson, Yale Obs., New Heaven, p. 401

\bibitem{}  Trager, S. C., Faber, S. M., Dressler, A., Oemler, A.,
1997, ApJ 485, 92

\bibitem{}  Wolfe, A. M., 1995, in {\sl QSO Absorption Lines},
ed. G. Meylan, Springer, p. 13

\bibitem{}  Wolfe, A. M., Prochaska, J. X., 1998, ApJ 494, L15

\bibitem{}  Woosley, S. E., Weaver, T. A., 1995, ApJS 101, 181

\bibitem{}  Worthey, G., Faber, S. M., Gonzalez, J. J., Burstein, D., 1994, ApJS 94, 687

\bibitem{}  Zaritsky, D., Kennicutt, R. C., Huchra, J. P., 1994, ApJ 420, 87

\bibitem{} van Zee, L., Salzer, J. J., Haynes, M. P., \etal, 1998, AJ 116, 2805

\bibitem{}  Zepf, S. E., Koo, D. C., 1989, ApJ 337, 34

\end{thebibliography}

}
\end{document}